\documentclass{egpubl}
\usepackage{eurovis2025}

\ConferencePaper        
\SpecialIssuePaper

\CGFccby

\usepackage[T1]{fontenc}
\usepackage{dfadobe}  

\usepackage{cite}  
\BibtexOrBiblatex
\electronicVersion
\PrintedOrElectronic
\ifpdf \usepackage[pdftex]{graphicx} \pdfcompresslevel=9
\else \usepackage[dvips]{graphicx} \fi

\usepackage{egweblnk}
\usepackage{xspace}

\usepackage{booktabs} 
\usepackage{tabularx}
\usepackage{subcaption}
\usepackage{enumitem}
\setlist[itemize]{leftmargin=*}

\newcommand{\ie}{i.e.{\xspace}}
\newcommand{\eg}{e.g.,\xspace}

\newcommand{\etc}{etc.{\xspace}}

\newcommand{\tool}{IntelliCircos\xspace}
\newcommand{\vshort}{circos plot\xspace}
\newcommand{\vshorts}{circos plots\xspace}

\newcommand{\User}{Bioinformatics analysts\xspace}
\newcommand{\user}{bioinformatics analysts\xspace}

\newcommand{\recommend}{\textit{Recom Edit Panel}\xspace}
\newcommand{\main}{\textit{Circos Dashboard}\xspace}
\newcommand{\config}{\textit{Configuration Panel}\xspace}
\newcommand{\reference}{\textit{Reference Panel}\xspace}
\newcommand{\data}{\textit{Data Panel}\xspace}
\newcommand{\actor}{Alex\xspace}

\newcommand{\rev}[1]{#1}

\usepackage{xcolor} 
\definecolor{REVISIONRED}{HTML}{DE3163}
\definecolor{REVISIONGREEN}{HTML}{9FE2BF}
\definecolor{REVISIONYELLOW}{HTML}{FFBF00}
\definecolor{REVISIONBLUE}{HTML}{6495ED}

\title[1143]%
      {\tool: A Data-driven and AI-powered Authoring Tool for Circos Plots}

\author[Mingyang Gu \& Jiamin Zhu \& Qipeng Wang \& Fengjie Wang \& Xiaolin Wen \& Yong Wang \& Min Zhu]
{\parbox{\textwidth}{\centering 
Mingyang Gu$^{1, *}$\orcid{0009-0008-5163-3058}, 
Jiamin Zhu$^{1, *}$, 
Qipeng Wang $^{1}$, 
Fengjie Wang $^{2}$, 
Xiaolin Wen$^{2}$, 
Yong Wang$^{2}$ and 
Min Zhu\thanks{Min Zhu (zhumin@scu.edu.cn) is the corresponding author.}$^{1}$
        }
        \\
{\parbox{\textwidth}{\centering 
$^1$Sichuan University, Chengdu, China\\
$^2$Nanyang Technological University, Singapore, Singapore\\
$^*$These authors contributed equally and should be regarded as joint first authors.
       }
}
}

\begin{document}


\maketitle
\begin{abstract}
Genomics data is essential in biological and medical domains, and \user often manually create circos plots
to analyze the data and extract valuable insights. 
However, creating circos plots is complex, as it requires careful design for multiple track attributes 
and positional relationships between them.
Typically, analysts often seek inspiration from existing circos plots, and they have to
iteratively adjust and refine the plot
to achieve a satisfactory final design, making the process both tedious and time-intensive.
To address these challenges, we propose IntelliCircos, an AI-powered interactive authoring tool that streamlines the process from initial visual design to the final implementation of circos plots. 
Specifically, we build a new dataset containing 4396 circos plots with corresponding annotations and configurations, 
which are extracted and labeled from published papers. 
With the dataset, we further identify track combination patterns, and utilize Large Language Model (LLM) to provide domain-specific design recommendations and configuration references to navigate the design of circos plots. 
We conduct a user study with 8 bioinformatics analysts to evaluate IntelliCircos, and the results demonstrate its usability and effectiveness in authoring circos plots.


\end{abstract}  
\section{Introduction}
Genomics data has played a vital role in biological and medical domains\rev{.}
\User typically use visualizations to analyze genomics data and gain valuable insights~\cite{nusrat2019tasks, o2021grand}.
Among various visualization techniques, \vshorts \cite{circos2024} are widely used for visualizing genomics data~\cite{o2018visualization}, which present gene data in various formats (\eg line, histogram) using multiple nested concentric circles arranged around the circumference of the same gene sequence.
Circos plots allow for a compact display of analysis results from multiple dimensions and facilitate multiple downstream tasks (\eg comparative analysis, correlation analysis) within a single visual space.

However, it is complex and time-consuming for \user to create \vshorts consisting of two main steps: \textbf{design} and \textbf{implementation}. 
During the design step, \user often refer to similar charts from other papers for inspiration, and then design the \vshort by considering various attributes, such as the track type, visual mapping for each track, as well as positional relationships between tracks.
Once the initial design is established, analysts typically use Circos-based programming tools (\eg Circos\cite{krzywinski2009circos}, TBTools~\cite{chen2020tbtools}) to implement the circos plot configuration 
based on their preliminary design, which requires iterative refinement and frequent update to achieve an optimal design. %
Furthermore, the authoring process often requires switching between tools for design and implementation.

Several tools have been developed to facilitate these two steps. 
Existing tools mainly focus on the implementation step.
\textit{Circos}\cite{krzywinski2009circos}, an authoring tool for implementing circos plots, provides various tracks and allows users to 
draw \vshort through the Perl language\cite{wall1994perl}.
Despite being widely used by \user, \textit{Circos} is not user-friendly (\eg steep learning curve, \rev{lack of interactivity in output charts}, and cumbersome configuration) \cite{cui2021interaccircos, naquin2014circus}.
As a result, several
tools have been 
developed to address implementation challenges with \textit{Circos}, including tools built in more accessible programming languages~\cite{zhang2013rcircos, sun2020rcirc, krzywinski2024circosjs}, tools enabling interactive analysis on the resulting plots (\eg zoom and click) on the created result~\cite{cui2021interaccircos, cui2016biocircos, cui2020ng, lyi2021gosling} 
and Graphic User Interface (GUI)-based tools that eliminate the need for coding and terminals~\cite{an2015j, yu2018shinycircos, wang2023shinycircos, wang2023enabling}. 
However, they still require intensive manual configuration from users, and users still face design challenges when using these tools.
ShinyCircos-V2.0~\cite{wang2023shinycircos} introduces a sample gallery to inspire users with \vshort designs, but it requires users to manually digest the samples first, then apply the learned knowledge to the subsequent design of \vshort, which is cognitively overwhelming.
To address this issue, GenoREC~\cite{pandey2022genorec} uses a rule-based approach to recommend genomics visualizations, assisting users in designing visualizations by considering the input data and tasks, simplifying the design step.
However, the predefined rules make it difficult to adapt to new tasks, thereby limiting its quality and applicability.
Also, it is difficult for \user to specify their design requirements using GenoREC, resulting in its poor usability.

To address these challenges, we propose \tool, a Large Language Model (LLM)-powered interactive authoring tool that streamlines the process from design to the actual implementation of circos plots.
First, we collect a dataset of \vshort illustrations and annotations from existing publications, where the quality of the \vshort is guaranteed. 
We then analyze the relationships of the common track combinations within this dataset. 
Next, we propose a \vshort recommendation method based on Retrieval Augmented Generation (RAG)~\cite{lewis2020retrieval}.
This method retrieves relevant examples from the above dataset based on user requirements, providing LLM with references and incorporating design patterns into prompts to improve recommendation performance in specific domain scenarios.
\tool allows users to browse and analyze configurations of similar charts, receive progressive design recommendations considering both user-specified constraints and current creation in the user interface, and supports easy refinement of the created circos plot.

We present a usage scenario to illustrate the usage of \tool, and further conducted a user study with 8 participants to evaluate its effectiveness and usability. The results demonstrate that \tool can effectively assist users in the design of \vshort and greatly simplifies the implementation of \vshort. The major contributions of this paper can be summarized as follows:

\begin{itemize}[nosep, left=1.8em]
    \item A dataset of \vshorts with extracted track combination patterns 
    to support and enhance the design recommendation process for \vshorts. 
    \item An LLM-powered and interactive \vshort authoring tool, \tool, which assists users in the designing and implementing circos plots.
    \item A user study to demonstrate the effectiveness and usability of \tool revealing its advantages and limitations. 
\end{itemize}

\section{Background} \label{sec:definition}
To enhance clarity, this section introduces several key terms frequently mentioned throughout the paper.

\begin{figure}[tb]
  \centering
  \includegraphics[width=0.95\columnwidth]{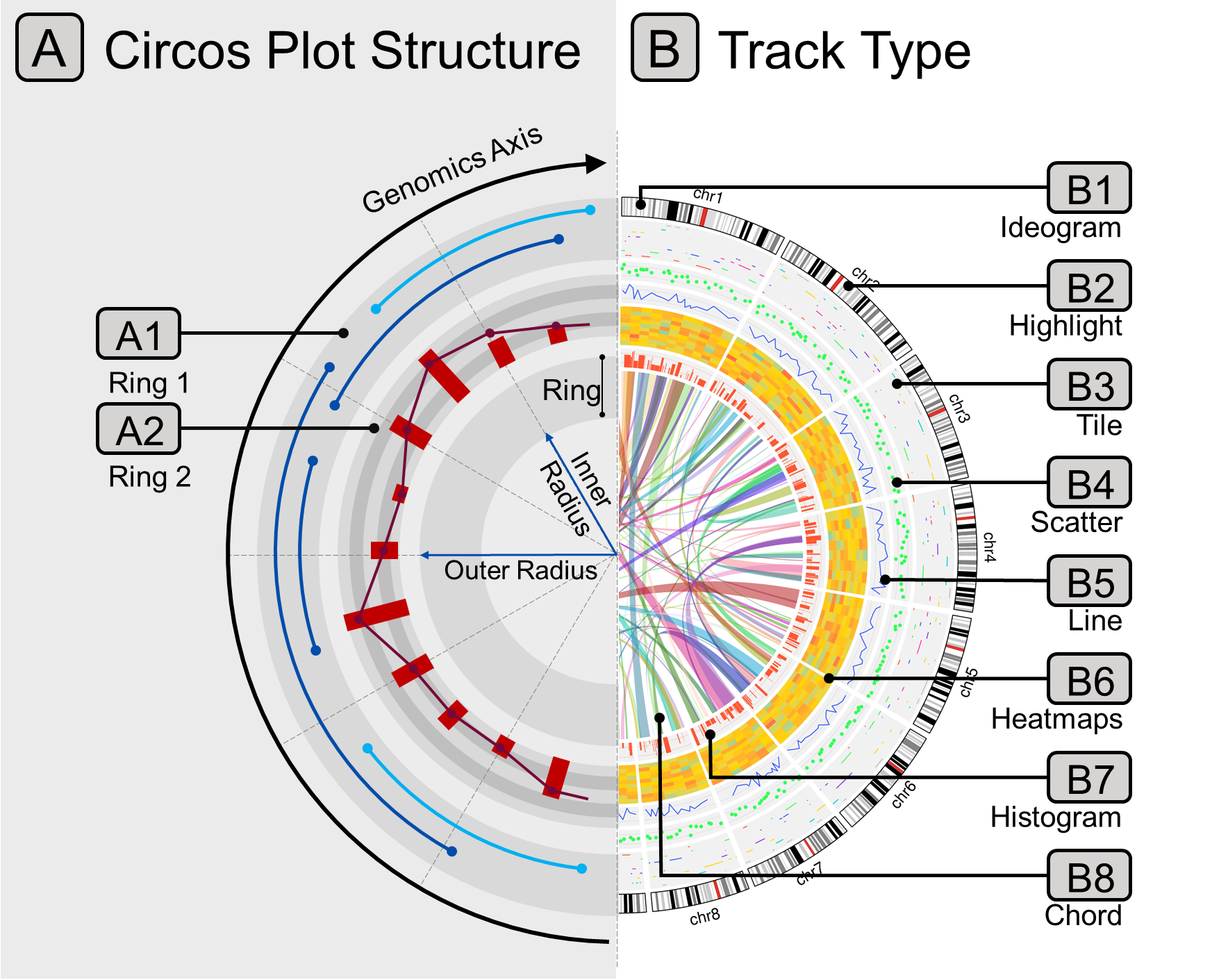} 
  \caption{
  (A) Structure of a Circos Plot.
  A circos plot consists of multiple nested \textbf{rings} surrounding a genomics axis, with each ring containing one or more \textbf{tracks}. 
  (B) Example of a Circos Plot.
  The tracks (B1-B8) are arranged from outermost to innermost.
  }
  \label{fig:background}
\end{figure}

Consistent with research on multi-view visualizations\cite{9222323, 9904451}, we define a \textbf{view} as a type of visualization that can be categorized into several common types\cite{borkin2013makes} (\eg line, histogram, and scatter, \etc).
In genomics, \textbf{track} is commonly used to refer to a view~\cite{9904451, nusrat2019tasks}, so 
we use the term ``track'' instead of ``view''
in the following discussion.
A track displays a single dataset~\cite{9904451}, and different tracks represent various aspects of genomics data~\cite{wang2023shinycircos}.
In addition to common visualization types, genomics tracks may include specialized visualizations, such as ideogram, highlight, and tile.
The position of a track is determined by its inner and outer radii: the inner radius refers to the distance from the inner side of the track to the circle’s center, while the outer radius refers to the distance from the outer side of the track to the center.
To enhance visual appeal or meet analytical needs, multiple tracks can be grouped within the same concentric rings, forming a structure we refer to as a \textbf{ring}.
For example, in \autoref{fig:background} (A), there are three nested rings, where the ring \autoref{fig:background} (A1) is composed of a tile, and the ring \autoref{fig:background} (A2) is composed of two tracks: line and histogram.
A \textbf{circos plot} typically consists of multiple concentric rings arranged around the same genomic axis.
Angular ranges are used to encode the position coordinates of data across all tracks, effectively mapping data’s location on the genomic axis.
For example, \autoref{fig:background} (B) shows a \vshort with eight tracks, listed from outer to inner as follows \autoref{fig:background} (B1 - B8): ideogram, highlight, tile, scatter, line, heatmaps, histogram, and chord. 
Circos plots are widely used to demonstrate similarities or differences among diverse genomic features associated with the same genomic regions\cite{wang2023shinycircos}.
Furthermore, Circos\cite{circos2024} is the most commonly used tool for creating circos plots.

In practice, tracks in circos plots are typically combined.
To better explore the relationships between tracks, we classify the track combinations into two categories: \textbf{stacked relationship} and \textbf{synthesized relationship}.
\rev{A stacked relationship occurs when two tracks are placed in rings without overlapping, while a synthesized relationship involves two tracks with overlapping inner and outer radii.}
For example, \textit{ideogram} (\autoref{fig:background} (B1)) and \textit{highlight} (\autoref{fig:background} (B2)) have a synthesized relationship, while their synthesized track and other tracks have a stacked relationship.

\section{Related Work} \label{sec:relatedwork}
The related work of this paper can be divided into two areas: circos plot authoring tools and visualization recommendation.

\subsection{Circos Plot Authoring Tools}
Circos plot authoring tools can be broadly classified into two types: \textbf{template-based configuration tools} and \textbf{interactive authoring tools}.
Template-based configuration tools, such as Vega-Lite\cite{satyanarayan2016vega}, ECharts\cite{li2018echarts}, and ggplot2\cite{wilkinson2011ggplot2}, pre-define data-to-visualization mapping rules, allowing users to input data and configure settings to create charts.
For circos plots, Circos\cite{krzywinski2009circos} was the first tool offering pre-built track templates and requiring users to write configuration files in Perl.
Subsequent tools like RCircos\cite{zhang2013rcircos}, ggbio\cite{yin2012ggbio}, circlize\cite{gu2014circlize}, and pyCircos\cite{ponnhide2024pycircos} have integrated circos plot creation into data analysis workflows using R and Python.
More recently, Gosling\cite{lyi2021gosling} enabled interactive genomics visualizations, including \vshort, via configuration. 
While these tools support the creation of circos plots, they often come with a steep learning curve\cite{wang2023enabling} and lack immediate visual feedback during the iterative configuration modification.

Interactive authoring tools, like J-Circos\cite{an2015j}, shinyCircos\cite{yu2018shinycircos}, and the circos plot module in TBtools\cite{chen2020tbtools}, simplify the encoding process by offering GUIs for interactively adjusting configurations.
However, they still require manual configuration.
AutoGosling\cite{wang2023enabling} improves on this by using an image detection model to generate corresponding Gosling configurations from user sketches or example images, but creating the sketches still requires additional user efforts.

Despite these advancements in bioinformatics visualization, circos plots, with their multiple tracks and complex configuration options, present significant design challenges for \user, especially those without the expertise of visualization.
So we can see that current tools lack design assistance, while also burdening them with operational complexities.
To address these gaps, we propose \tool, a tool designed to assist \user in creating circos plots through visualization recommendations, design references and interactive configuration.

\subsection{Visualization Recommendation}
Recommending precise and effective data visualizations is a challenging task\cite{qin2020making}.
Various visualization recommendation methods have been proposed to simplify this process, typically falling into two categories: rule-based and machine learning (ML)-based methods\cite{zhu2020survey}.
Rule-based methods rely on expert-defined heuristic rules to recommend visualizations\cite{mackinlay2007show, wongsuphasawat2015voyager, wongsuphasawat2017voyager}.
For example, GenoREC\cite{pandey2022genorec} recommends genomics visualizations based on the user-uploaded data and selected parameters.
However, its rigid rules limit the types of data and tasks it can accommodate, making it challenging to adapt to diverse scenarios.
Additionally, developing comprehensive rules for recommending \vshorts is a challenging task\cite{10254497}.

In contrast, ML-based methods are data-driven approaches that allow algorithms to autonomously learn visualization design principles from high-quality examples.
These methods have shown promising performance\cite{pandey2022genorec}.
Draco\cite{moritz2018formalizing} uses answer set programming to generate visualizations while learning soft constraints for design.
DeepEye\cite{luo2018deepeye} rank visualizations based on input data, tasks, and context.
VizML\cite{hu2019vizml} uses more than one million dataset-visualization pairs to recommend design choices.
Recent studies have also explored the use of Large Language Models (LLMs) for visualization recommendations, including Chat2VIS\cite{maddigan2023chat2vis} and LIDA\cite{dibia2023lida}.
Besides, ChartGPT\cite{tian2024chartgpt} creates a dataset consisting of abstract utterances and charts and improve model performance through fine-tuning, transforming the process of generating charts from natural language into a step-by-step reasoning workflow.
The Prompt4Vis framework\cite{li2024prompt4vis}, enhances the model's generation capabilities by providing external information through a database, leverages LLMs and context learning to improve the performance of generating data visualizations from natural language.

ML-based recommendation methods typically rely on high-quality datasets to achieve good performance. 
However, current recommendation methods mainly focus on general domains and are not suitable for recommending circos plots. 
Therefore, we developed an ML-based recommendation method for circos plots, while simultaneously collecting a \vshort dataset from published papers to power design recommendations, aiming to provide more practical and tailored guidance for circos plot design.

\section{Design Goals} \label{sec:design goals}

Our tool is designed to assist \user in creating circos plots by addressing both design and implementation aspects. 
Based on relevant literature\cite{wang2023enabling, pandey2022genorec, wang2023shinycircos},
we derive the following design goals.

\textbf{G1: Assist in the design of circos plots.}
Currently, support for designing \vshorts remains limited, as evidenced in the literature. 
However, the design stage is inherently challenging, requiring consideration of factors such as track types, visual mappings for each track, and positional relationships between tracks. 
Furthermore, \user usually lack expertise in visualization, which exacerbates the difficulty of the task. 
These challenges highlight the need for a streamlined approach to simplify the \vshort design stage and better support users in their creation efforts.

\begin{itemize}[nosep, left=1.8em]
\item 
\textbf{G1.1: Enable users to describe their design requirements in a natural way.}
Current tools do not allow users to easily express their design requirements. 
Additionally, \user often lack specialized visualization design experience and may only have vague needs \cite{vaithilingam2024dynavisdynamicallysynthesizedui}, while advanced natural language interfaces can effectively improve the usability of visualization tools\cite{yu2019flowsense, srinivasan2017natural, shen2022towards}. 
Therefore, our tool should support users in naturally expressing their design needs.

\item 
\textbf{G1.2: Support progressive design recommendation.} 
The design of \vshort is an iterative process, which is influenced by users' requirements, data, downstream analysis tasks, and existing design patterns \cite{todi2016user}.
Moreover, users' needs may vary throughout the design stage.
For example, the track layout is considered at the initial design stage, while fine adjustments (\eg color, transparency) are made after the layout is completed.
As a result, relying on recommendation to generate a \vshort entirely from scratch is often impractical.
Thus, the tool should provide recommendations in a progressive manner: offering incremental recommendations based on users' current needs and the current design \cite{Long2022}.

\item 
\textbf{G1.3: Support browsing and analyzing visual configurations of similar circos plots.} 
A common process for creating \vshorts is to search for similar papers and examine the corresponding figures in those papers\cite{wang2023shinycircos}.
Then, \user seek design inspiration coordinates these illustration configurations.
Therefore, the tool should support the process of finding similar \vshorts and providing their configurations.

\end{itemize}

\textbf{G2: Simplify the implementation of \vshort.}
\User often need to manually implement designs using the Circos software, a process that is time-consuming and lacks immediate visual feedback of modifications.
Additionally, for those unfamiliar with Circos, the learning curve is steep.
Therefore, an intelligent, interactive authoring tool is needed to streamline the design process and enable more efficient implementation.

\begin{itemize}[nosep, left=1.8em]
\item 
\textbf{G2.1: Automatically apply the selected design.} 
Currently, users must manually implement designs in \textit{Circos} to assess their effects via programming, which is tedious and time-consuming.
Besides, \textit{Circos} \rev{does} not provide an intuitive, real-time display of configuration changes, which forces \user to repeatedly trial and error to achieve the desired effect \cite{tyagi2022infographics}.
Therefore, the tool should automatically apply selected designs and update the Circos plot in real time to streamline the design and implementation process when creating \vshorts \cite{shi2023destiji}.

\item 
\textbf{G2.2: Support easy refinement of track configurations.}
Allowing users to exert control over the \vshort creation process can serve as a way of refining the recommended design to better meet users' individual preferences and needs\cite{heer2019agency, myers2022human}, and mitigate the limitations of automated methods\cite{blohm2018comparing, li2023ai, zukerman2001natural}.
Therefore, the tool should support the refinement of the recommended results with a suite of easy-to-use interactions, such as adding or deleting tracks, matching tracks to data and adjusting various track configurations.

\end{itemize}

\section{Dataset}
To support our recommendation method, we built a circos plot dataset including three modalities: circos plots (bitmaps), annotations (natural language), and configurations (structured files).
We constructed the dataset through a three-step workflow: paper collection, illustration-annotation pair extraction, and illustration labeling. We then conducted a track combination analysis to identify common patterns.
The workflow and analysis details are as follows:

\textbf{Paper Collection.}
To ensure our dataset aligns with genomic themes and includes professionally created circos plots, we collected circos plot illustrations from bioinformatics published papers.
Specifically, we downloaded 3,377 PDF (Portable Document Format) documents from SemanticScholar\cite{semanticscholar} and Unpaywall\cite{unpaywall} that 
cited  \textit{Circos: An information aesthetic for comparative genomics}\cite{krzywinski2009circos}, a paper widely cited by studies utilizing \textit{Circos} for circos plot creation.

\begin{figure}[tb]
  \centering
  \includegraphics[width=\columnwidth]{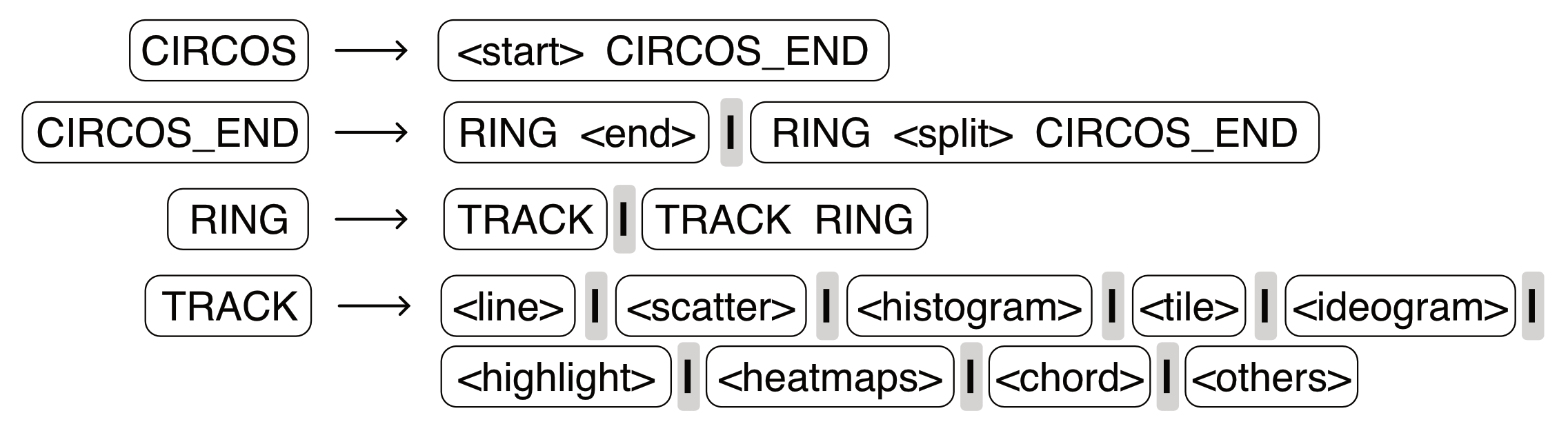} 
  \vspace{-6mm}
  \caption{
  \rev{
    Label syntax of a circos plot configuration. 
    A \textbf{CIRCOS} begins with the <start> terminator and ends with \textbf{CIRCOS\_END}, containing one or multiple \textbf{RING} elements.
    A \textbf{RING} consists of \textbf{TRACK} elements, and \textbf{RINGs} are separated by the <split> terminator, ordered from outside to inside. 
    A \textbf{TRACK} represents a single track within a \vshort.
    We classify text tracks, scale tracks, and other non-visual tracks under <others>, as our focus is on the design of visual elements.
    }
    }
  \label{fig:CTML}
\end{figure}

\textbf{Illustration-Annotation Pair Extraction.}
To automate the manual extraction of circos plot illustrations and their corresponding annotations from numerous papers, we developed a deep learning-based method to identify illustration-annotation pairs and filter out irrelevant information.
We used PDFFigure2.0\cite{clark2016pdffigures} to render PDFs into bitmaps and PyMuPDF\cite{pymupdf} to locate and crop illustrations, excluding images smaller than 10\% of the page width (\eg icons).
Poppler's paragraph grouping mechanism\cite{poppler} was then applied to match annotations with illustrations, resulting in 29,701 matched pairs with metadata such as paper source, page number, and position.
To isolate circos plot illustrations, we trained a binary classification deep learning model using ResNet50\cite{koonce2021resnet} fine-tuned with pre-trained weights from IMAGENET1K\_V2\cite{deng2009imagenet}.
Positive samples were sourced from the \textit{Circos}\cite{circos2024} website and image searches on Google and Bing, while negative samples came from VIS30K\cite{chen2021vis30k}.
All images were in PNG (Portable Network Graphics) format and resized to 224*224 pixels as input.
The dataset was split into 80\% for training and 20\% for testing. 
After training for 5 epochs, the model achieved 99.23\% accuracy on the test set.
Using this model, we extracted 4,531 illustration-annotation pairs. After manual verification for accuracy and completeness, we retained 4,396 valid pairs.

\begin{figure*}[tb]
  \centering
  \includegraphics[width=\textwidth]{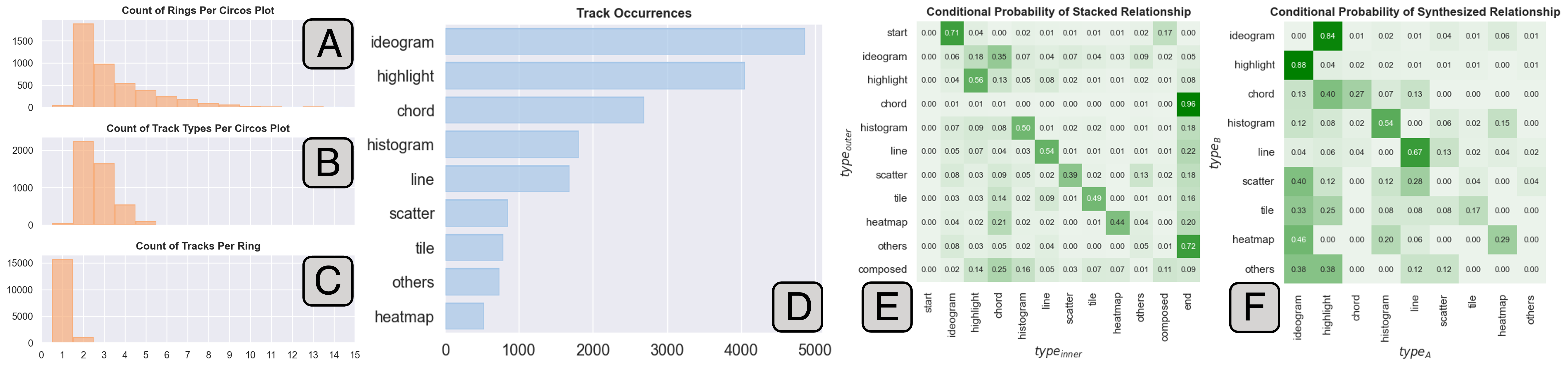} 
  \vspace{-6mm}
  \caption{ Track Combination Analysis Results: 
  (A) The number of rings in each \vshort,
  (B) The number of track types in each \vshort, 
  (C) The number of tracks in each ring, 
  (D) The number of tracks of each type, 
  (E) The conditional probability of stacked relationship, 
  and (F) The conditional probability of synthesized relationship. 
  } 
  \label{fig:statistic}
  \vspace{-4mm}
\end{figure*}

\textbf{Illustration Labeling.}
Inspired by the \textit{Circos} configuration format and XML (Extensible Markup Language) syntax,
we use the syntax in \autoref{fig:CTML} to label illustrations.
\rev{
This syntax offers a structured representation to capture the core elements of circos plots, \ie, structures and track types.
While it conceals implementation details to minimize the burden and cost of dataset annotation.
}
To facilitate this process, we developed a web-based labeling tool and enlisted five annotators with visualization experience. 
Each \vshort was labeled by two annotators. 
On average, each annotator labeled approximately 1,800 illustrations, and the entire process took six weeks.

\textbf{Track Combination Analysis.}
We conducted a detailed analysis of the track combinations used in circos plots within our dataset.
First, we examined the number of rings (\autoref{fig:statistic} (A)) and tracks types (\autoref{fig:statistic} (B)) per plot.
The results show that many plots use more than five rings, indicating complex requirements for track combination and ordering.
Next, we analyzed the number of tracks per ring (\autoref{fig:statistic} (C)), revealing that multiple tracks are typically stacked rather than synthesized within the same ring.
\autoref{fig:statistic} (D) illustrates the distribution of different tracks in the dataset.
To further understand the relationships between track types, we calculated pairwise relationships for the two primary track combination modes: stacked and synthesized.

Tracks in stacked relationships are displayed in a linear order. 
We used conditional probability to calculate the pairwise relationships between adjacent tracks ($type_{inner}$ and $type_{outer}$):
\begin{equation}
P(type_{inner}|type_{outer}) = \frac{P(type_{outer} \cap type_{inner})}{P(type_{outer})},
\end{equation}
where $P(type_{outer}|type_{inner})$ ranges from [0, 1]; values close to 0 indicate that $type_{inner}$ rarely appears adjacent to $type_{outer}$, while values close to 1 indicate frequent adjacency.
Additionally, we designated tracks combined in synthesized relationships as \textit{synthesized tracks}, and involved the \textit{start track} (located outside the outermost track) and the \textit{end track} (located inside the innermost track) to analyze the track types at the boundaries.
The results (\autoref{fig:statistic} (E)) reveal that many track types like histogram, line, scatter, and heatmaps have probabilities above 0.5 (diagonal cells), suggesting that circos plots tend to stack multiple tracks of the same type.
Furthermore, the high proportion of chord and other views in the \textit{end track} column indicates that the central position is often reserved for displaying chord or other customized, user-defined views.

For tracks combined in synthesized relationships, we calculated the conditional probability of two track types ($type_{A}$ and $type_{B}$) appearing together:
\begin{equation}
P(type_{B}|type_{A}) = \frac{P(type_{A} \cap type_{B})}{P(type_{A})}.
\end{equation}
The results (\autoref{fig:statistic} (F)) indicate that the ideogram and highlight exhibit the strongest connection, as they are frequently combined within the same track. 
This is likely because ideograms, which display chromosomes and bands, enhance the interpretability of gene regions highlighted in the data.
Additionally, histogram and line are often paired with themselves to present multiple data attributes, likely due to their ability to use colors to distinguish different attributes, making them ideal for visualizing complex datasets.

\section{\tool}
Based on the collected dataset, we developed a \vshort recommendation framework, configuration reference and an authoring tool for \vshort creation. 
The recommendation framework and configuration reference together provide design assistance to users when creating circos plots with \tool.

\subsection{Recommendation Framework}

\begin{figure}[tb]
  \centering
  \includegraphics[width=\columnwidth]{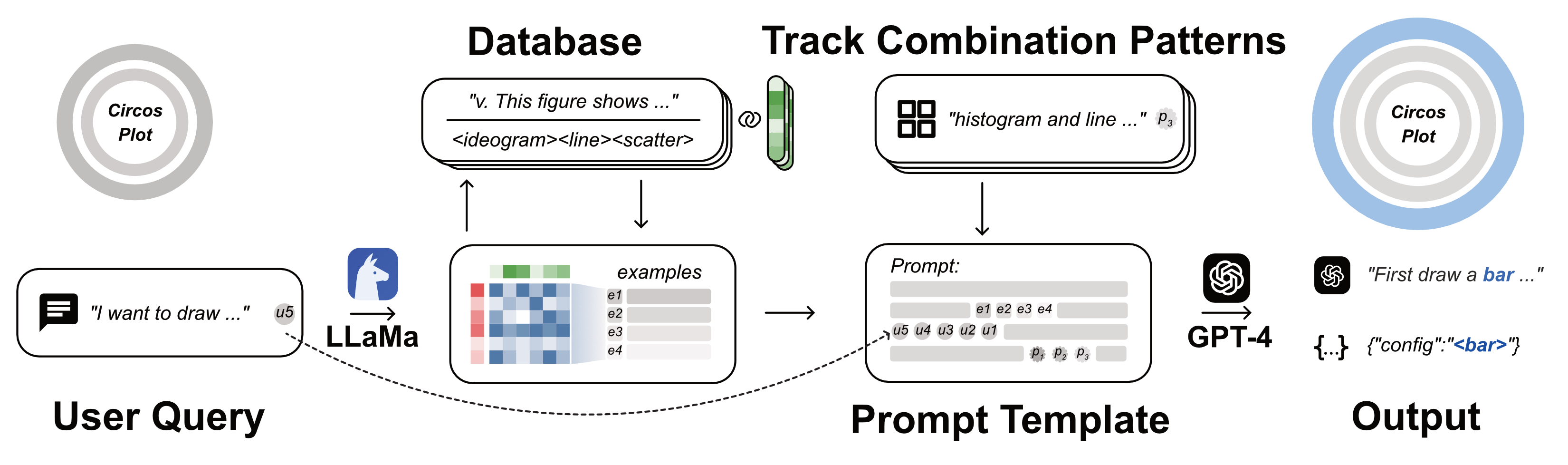} 
  \vspace{-6mm}
  \caption{
  Recommendation Workflow.
  For the user query, the tool encodes it into a vector using LLaMa and searches for semantically similar samples in the database.
  The retrieved samples, user query, and the results from the track combination analysis in Section 4, are combined to form the prompt for GPT-4, which generates the recommendation results.
  }
  \label{fig:recommend}
\end{figure}

Our recommended method uses a Retrieval Augmented Generation (RAG)\cite{lewis2020retrieval} framework, 
a widely adopted approach that combines a model's parameters with external non-parametric memory to generate language, thereby overcoming the limitations of language models in domain-specific tasks.
RAG consists of two stages: a retriever and a generator. 
The retriever takes the user's query and maps both the query and documents from an external corpus into a high-dimensional vector space using a shared embedding model. By calculating similarities (e.g., Euclidean distance) between the query vector and document vectors, it identifies the most relevant documents.
The generator, often a pre-trained large language model (LLM), then uses the content of these retrieved documents to produce natural language text that satisfies both syntactic and semantic requirements. 
The prompt input to the generator is optimized through prompt engineering to guide the model in generating the most relevant output.
Our recommendation framework is also divided into two stages based on RAG: the similar sample retriever and the configuration generator, as shown in \autoref{fig:recommend}.

\textbf{Similar Sample Retriever.} 
We developed the Similar Sample Retriever to search for samples that are related to \user's query from the collected \vshort illustration-annotation pairs. 
The Similar Sample Retriever consists of two components: a text embedding model and a circos plot vector database which stores encoded \vshort annotations and corresponding configuration vectors.
First, we selected the LLM-Embedder\cite{zhang2023retrieve} as the base model for calculating text embeddings, as it is optimized for retrieval-augmented tasks with LLMs.
The text embedding model converts a user's query into a 1024-dimensional embedding vector.
Similarly, it converts each annotation and its corresponding configuration in the dataset into vectors within the same semantic space.
Then, we use Chroma\cite{chroma-docs} to build the vector database, which stores annotation-configuration vectors generated by the text embedding model, providing external knowledge for the subsequent generator.
The vector converted from the user's query is compared with vectors stored in the database using Euclidean Distance, and the top 10 most relevant vectors are selected as examples, where the number of examples can be adjusted by users.
Since both the annotations and the query are embedded in the same semantic space, the similarity between vectors reflects their semantic similarity.

\textbf{Configuration Generator.}
We then created the Configuration Generator based on the GPT-4 to recommend circos plot configurations.
The Configuration Generator uses a prompt template to guide the model's responses during the recommendation process.

System prompts input for the model at the start of dialogue include:
\textbf{Task introduction} succinctly describes the model's role and responsibilities: \textit{``You are an expert of Genomics Visualization. Your job is to recommend Circos plot configurations that meet users' needs.''}
\textbf{Background knowledge} covers the visual structure of a \vshort, track types and configuration format.
Constraints are included to enforce format restrictions, such as
\textit{``You can only use the tokens listed below, and you cannot use any custom tokens''}. 
Additionally, guidelines like \textit{``If a ring includes many tracks, it is usually placed in the front of the sequence (i.e., the outer side) to avoid visual confusion.''} help the model understand common track combination patterns in \vshorts.

Once a user sends a query, the prompts are processed and passed to the model:
\textbf{Examples} consist of sample visualizations retrieved by the Similar Samples Retriever, each example containing annotations and configurations. 
\textbf{Requirements} capture the user’s natural language input.
\textbf{Existing design} refers to the current configuration created by users, if available.

We standardized the model's expected input and output through prompts to ensure consistency. 
Finally, the output from the Configuration Generator is used to recommend circos plot configurations, and we extract the track configuration from this result.

\subsection{Configuration Reference} \label{sec:reference}

In addition to recommending circos plots, browsing and analyzing similar \vshort configurations from existing papers is essential for the design stage. 
To better help user retrieve configurations, we combine two retrieval modes.
The first mode retrieves configurations based on the latest user's query, selecting similar configuration results by Similar Sample Retriever. 
The second mode retrieves configurations based on the Levenshtein distance between the current design configuration.
The Levenshtein distance is a metric that measures the difference between two sequences by counting the minimum number of single-character edits (insertions, deletions, or substitutions) required to transform one string into the other.
Both retrieval modes retrieve the top 10 most similar configurations, with users able to adjust the number of retrieved configurations.

Initially, the retrieved configurations are displayed in text form, which is unintuitive.
We convert the configuration set into a Directed Acyclic Graph (DAG).
\rev{
The nodes of the graph represent the track types in the configurations, along with \textit{\textless start\textgreater}, \textit{\textless end\textgreater}, and the separator \textit{\textless split\textgreater} nodes. 
Nodes within the same configuration are connected by edges. 
Since the DAG contains redundant information, we optimize its structure by merging nodes of the same type along with their edges. 
The number of merged edges is accumulated as the edge weight.
}

\subsection{User Interface}

\begin{figure*}[tb]
 \centering 
 \includegraphics[width=0.95\linewidth]{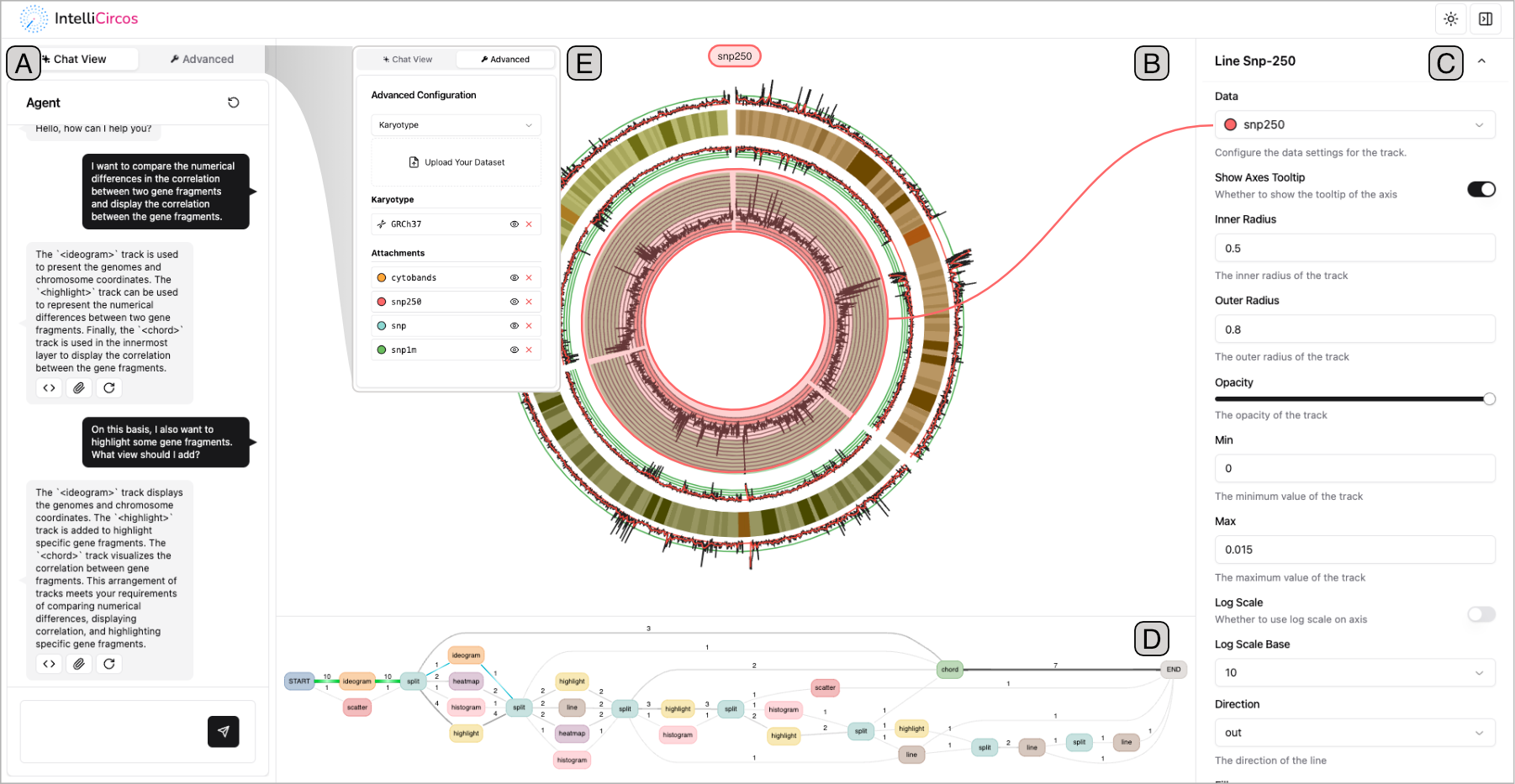}
 \caption{ 
  \tool is an interactive AI-powered authoring tool designed to facilitate the creation of \vshort.
  Its user interface comprises five interconnected components: 
  \recommend (A) generates design recommendations based on user needs in natural language,
  \main (B) renders the \vshort created by users,
  \config (C) offers easy refinement of track configurations,
  \reference (D) enables users to browse and analyze configurations of similar \vshorts,
  \data (E) allows users to manage the source data.
 }
 \label{fig:ui}
\end{figure*}

This section presents five interactive modules in the user interface to support users in creating \vshorts.

\textbf{Circos Dashboard} (\autoref{fig:ui} (B)) shows a \vshort the user is currently creating, offering a quick preview of the plot. 
When the user hovers over any track, \tool highlights the selected track and guiding lines will connect it to the corresponding \data and \config.
\rev{
Since each track type is associated with a standardized data format, 
we adopt the mechanism from CircosJS \cite{krzywinski2024circosjs} to map the data to the corresponding chart.}

\textbf{Recommend Panel} (\autoref{fig:ui} (A)) implements interactive \vshort configuration recommendations.
Users can describe their design requirements in the input box, which offers an auto-complete feature to enhance input, where \tool enhances input efficiency with an auto-complete feature.
Command templates are also available: typing ``\textbackslash recommend'' triggers suggestion for track combinations based on the current configuration, and ``\textbackslash data'' adds the all dataset name and type (karyotype data and attachment data) to the start of the user input.
The conversation history with the recommendation model is displayed in dialogue bubbles.
Each model's output bubble contains recommendation results, relevant papers, and a regenerate button.
Users can click on the recommendation result to preview and apply it to the current configuration.

\rev {
\textbf{Reference Panel} (\autoref{fig:ui} (D)) visualizes the constructed DAG from \autoref{sec:reference} using a node-link flow diagram.
We apply the Sugiyama algorithm \cite{nikolov2016sugiyama} to optimize the DAG layout, minimizing edge crossings while showing its hierarchical structure.
A path from \textit{\textless start\textgreater} to \textit{\textless end\textgreater} represents a circos plot configuration. 
Nodes are color-coded based on their types to enhance visual distinction. 
Edges connect nodes, with edge width indicating the number of paths passing through them, and the exact count displayed above each edge.
To aid interpretation, 
edges in the user's current configuration are colored green, 
edges in the recommended path from the \recommend are blue, 
and all other edges are gray.
Users can hover over a node to highlight all paths passing through it, revealing common configuration patterns.
Clicking on a node allows users to create a circos plot with ease in the \main.
\tool will construct a path from the \textit{\textless start\textgreater} to the selected node to form the configuration for the new plot. 
When constructing the path, the tool preserves segments from the user's current configuration. 
For the remaining parts, \tool prioritizes paths recommended by the \recommend. 
If no match is found, it selects the most frequently occurring path.
}

\textbf{Configuration Panel} (\autoref{fig:ui} (C)) allows users to manually insert, edit, and delete tracks.
The global control form is used to create new tracks, while each track has a corresponding local configuration form with customizable settings such as color and scale. 
\tool simplifies the process to reduce users' workload by automatically selecting initial data sources and configuration parameters based on heuristic rules, such as prioritizing unused data sources and copying configurations from tracks of the same type.
Any modifications made to the configuration are displayed in real-time on the Circos Dashboard.

\textbf{Data Panel} (\autoref{fig:ui} (E)) allows users to manage data for their circos plots.
It supports data upload, preview, and deletion.
Similar to \textit{Circos}' data management, data is categorized into \textit{karyotype data}, which marks chromosome and gene hierarchy information, and \textit{attachment data}, which marks quantitative/qualitative data associated with genes.
Users can set unique color markers for each type of source, distinguishing them visually in the display area.

\section{Usage Scenario}

\begin{figure}[tb]
  \centering
  \includegraphics[width=\columnwidth]{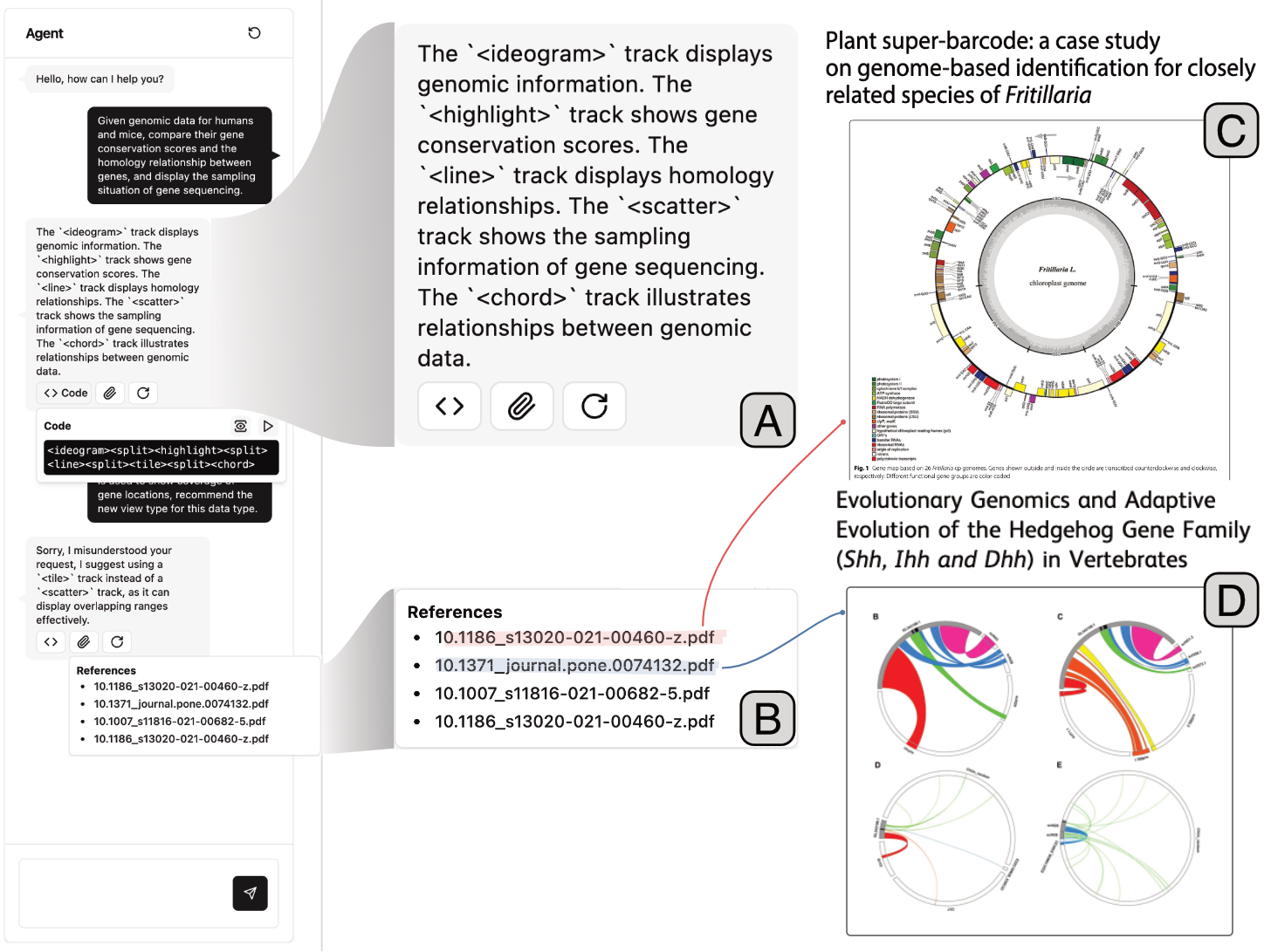} 
  \vspace{-7mm}
  \caption{ This figure illustrates a usage scenario in which a bioinformatics analyst, \actor, utilized \recommend to create \vshort,
  following 
  (A) gain a design recommendation,
  (B) find reference papers that are similar to the requirements,
  and (C-D) check reference papers' corresponding illustration and annotation.
  } 
  \label{fig:case}
    \vspace{-4mm}
\end{figure}

Here we present a scenario to illustrate how \tool supports the design and implementation of circos plots.

Suppose, \actor, a genomics researcher, planned to use \tool to design a \vshort comparing human and mouse genomes in terms of gene conservation scores, homologous relationships, and gene sequencing sampling.

\actor uploaded karyotype datasets that describe the correspondence between gene locations and chromosomes \rev{for both humans} and mice, along with attachment datasets via the \data.
\tool parsed and displayed these files.

Then, \actor started to create a \vshort. 
Firstly, \actor opened the \recommend to request recommendations with a prompt for design inspiration:
``Given genomic data for humans and mice, compare their gene conversation scores and the homology relationship between genes, and display the sampling situation of gene sequencing.''
\tool provided the following recommendation: <ideogram><split><highlight><split><line><split><scatter>
<split><chord>.
This can be interpreted as: the <ideogram> track displays genomic information, the <highlight> track shows gene conservation scores, the <line> track displays homology relationships, the <scatter> track shows the sampling information of gene sequencing, the <chord> track illustrates relationships between genomic data (\autoref{fig:case} (A)). 
However, \actor felt the <scatter> track was improper for displaying gene sequencing sampling. 
She re-entered the prompt, requesting a new track: ``The sampling profile refers to gene segments measured in high-throughput sequencing and is used to show coverage of gene locations, recommend the new view type for this data type.'' 
Subsequently, \tool recommended: <ideogram><split><highlight><split><line><split><tile><split>
<chord> and explained: ``I suggest using a <tile> track instead of a <scatter> track as it can display overlapping ranges effectively.''

\actor was satisfied with the recommendation but wanted to further understand the rationale behind the decision.
She first clicked the References button to view the paper cited by the generated results (\autoref{fig:case} (B)).
One \rev{file, named} ``10.1186\_s13020-021-00460-z'', included an illustration with two stacked heatmaps representing the values of gene segments with different meanings. 
This aligned with the requirement described in \actor's initial prompt for comparing gene conservation scores (\autoref{fig:case} (C)).
Additionally, \actor noticed that the figure legend in the paper \rev{ file, named} ``10.1371\_journal.pone.0074132'', mentioned ``homology'' (\autoref{fig:case} (D)), which aligns with \actor's request for displaying homologous relationships in the prompt.
\actor appreciated the relevance of the references and the quality of the recommendation.

Then, \actor applied the recommendation results to the \main.
She found the <line> track less readable for presenting homology scores, so she explored alternatives in the \reference. 
Since the <histogram> track was highly recommended, she deleted the line track and clicked the histogram node to directly apply the design to the \main.

Additionally, \actor noticed that the default color scheme used the same color for both human and mouse chromosomes, which caused visual confusion.
To solve this problem, she opened the color selection panel corresponding to the ideogram track in \config and used the tool's preset color combinations to distinguish these species.
Finally, she felt that the height and direction of the histogram did not align with her preference, so she modified these settings in \config.
Now, \actor was satisfied with the final design, and exported it as an SVG (Scalable Vector Graphics) file to share with other colleagues.

\section{User Study}
To evaluate the usability and effectiveness of \tool, we conducted a user study with 8 participants ($\mathrm{P_1}$-$\mathrm{P_8}$).
They were asked to use \tool to create circos plots and give both qualitative and quantitative feedback.
The study setting is introduced in Section 8.1, and the results are presented in Section 8.2. 
\rev{
We did not conduct a comparative study since there are no appropriate baselines.
Existing tools like \textit{Circos} and \textit{RCircos} are often used by \user. Also, these tools primarily assist in the implementation while overlooking the design process.
In contrast, \tool supports both design and implementation, seamlessly integrating these two stages. 
It makes it difficult and unfair to compare \tool with existing tools.
As a result, we opted for a qualitative assessment of users' experiences with \tool, comparing it to their current practices.
}

\subsection{Study Design} \label{sec:study design}
We introduce the setting of the study here, including: participants, task, data, and procedure, as outlined below.

\textbf{Participants.}
\rev{We recruited eight graduate students from different local universities in biology-related majors,}
each with over three years of experience in circos plots creation.
All participants are familiar with creating circos plots, with a frequency of 1-5 plots per month.
Participants self-reported their familiarity with \textit{Circos} with a rating of 5.9 ± 1.1, where 1 represents ``No Experience'' and 7 represents ``Expert''.

\textbf{Task and Data.}
In our user study, participants were asked to create a circos plot using \tool.
They were required to use the provided data as sufficient as possible.
\rev{We utilized example data from CircosJS \cite{krzywinski2024circosjs}, which includes track data for plotting alongside chromosomes, label data for annotation elements in different tracks, and link data to establish connections between pairs of genomic regions.
This dataset meets the requirements of circos plots, including relationships and multi-layered annotations across various scales \cite{circos2024}, and its manageable size ensured the study did not
last for too long.}

\textbf{Procedure.}
All studies were conducted through one-to-one in-person meetings, each lasting about an hour.
Prior to the user study, we briefly introduced the study procedure to participants and gained their consent for video recording the whole process.
The evaluation consisted of four stages: a training session, an experimental session, a post-study questionnaire session, and a post-study semi-structured interview.
During the training session, we first introduced the features of \tool, and demonstrated the whole usage process by creating an example \vshort(10 minutes).
Then, participants were asked to freely explore the system to familiarize themselves with the features (5 minutes).
Next, they were asked to create a circos plot using \tool (25 minutes).
After this, participants were asked to complete a 7-point Likert questionnaire in a think-aloud protocol, where users were encouraged to justify their ratings (5 minutes). 
Specifically, a rating of 1 represents ``strongly disagree'', while a rating of 7 represents ``strongly agree''.
Finally, the study concluded with a semi-structured interview (15 minutes).
This interview included topics on the practices with exiting tools, and experience with \tool.

\subsection{Results} \label{sec:study result}
\begin{figure*}[tb]
 \centering 
 \includegraphics[width=\linewidth]{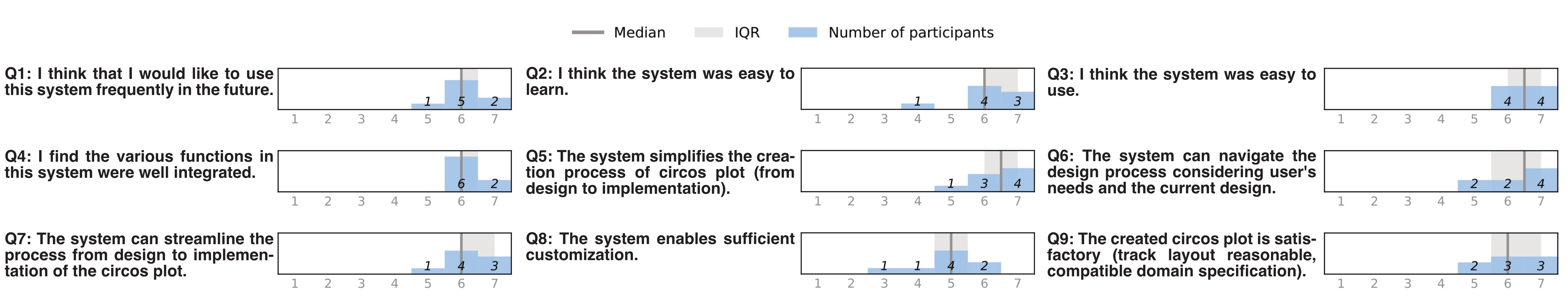}
  \vspace{-6mm}
 \caption{Questionnaire results. We designed our questionnaire in two aspects: the tool's usability (Q1 - Q4) and effectiveness (Q5 - Q9). Experts are asked to rate by a 7-point Likert scale (from 1 - strongly disagree to 7 - strongly agree).
 }
 \label{fig:questionnaire}
 \vspace{-4mm}
\end{figure*}

In the following, we report our user study results including challenges of current circos plot creation tools, participants' questionnaire ratings (\autoref{fig:questionnaire}), and feedback to our tool.

\textbf{Challenges of Current Circos Plot Software.}
All participants reported that \textit{Circos} is the mostly used tool to create circos plots.
The process of creating a circos plot with \textit{Circos} can be roughly divided into four stages: 
(a) searching for similar circos plot designs in literature for reference, 
(b) designing the sequence of tracks and their attributes (\eg view type, color) using formats like slides or sketches,
(c) implementing the design in \textit{Circos},
(d) repeating steps a-c until a satisfactory circos plot is achieved. 
They agreed that using \textit{Circos} to create a circos plot had the following problems.
\rev{
First, \textit{Circos} lacks design recommendations tailored to users' specific requirements or references for similar needs, making stages (a) and (b) both challenging and time-consuming.
}
Second, implementing designs through configuration is cumbersome, particularly for novice designers, as \textit{Circos} has a steep learning curve.
Third, \textit{Circos} does not properly support the iterative process of creating circos plots, leading to frequent trial and error. 
As $\mathrm{P_2}$ mentioned, \textit{``While the process of using Circos can gradually optimize my design, it’s too cumbersome.''}

The results confirm our hypothesis that current practice of creating circos plots has significant shortcomings.
It also validates our design goals and the development of \tool.

\textbf{Questionnaire Responses.}
The quantitative results of our user study reflect participants’ ratings on both the usability and effectiveness of \tool.

In terms of usability, \autoref{fig:questionnaire} (Q1 - Q4) depicts the distributions of ratings, as well as the medians (MDs) and interquartile ranges (IQRs) for all participants.
Overall, \tool was easy to learn and use, with both aspects receiving median ratings of 6 or higher.
While some participants gave low ratings on certain aspects, all expressed interest in using \tool in their future work (Q1).
For instance, $\mathrm{P_6}$ gave \rev{scores under four} for Q8 but still rated 6 for Q1.
$\mathrm{P_5}$ rated 4 for Q2, stating the steep learning curve for novice genomics researchers.
Notably, during the interview, all participants praised the seamless integration between design and implementation, as well as the three interactive modules (\recommend, \reference, and \config).

The distribution of ratings for effectiveness is shown in \autoref{fig:questionnaire} (Q5 - Q9).
Overall, most participants were satisfied with \tool.
Specifically, 7 out of 8 participants appreciated how \tool simplified the creation process of circos plots (Q5), with a median rating of 6.5 (\textit{IQR} = 1).
They also praised the quality of the final circos plot (Q9) and found design recommendation (Q6) and the design-to-implementation process (Q7) useful and aligned with their expectations.
However, the rating for customization support (Q8) was relatively lower, with most participants providing more neutral scores and one participant, $\mathrm{P_6}$, giving a negative score of 3, stating that \textit{``
The tool does not support customization of the chart or legend in the center.
''}
This reflects \tool's limited editing capabilities compared to mature software (e.g., \textit{Circos}).

\textbf{Feedback on \tool.}
Overall, feedback from all participants on \tool was positive.
Participants found the user interface of \tool intuitive and well-organized, and expressed a strong willingness to use our tool.
Below, we group their feedback according to the design goals outlined in Section 4.

G1: Assist in the design of circos plots.
When asked about \tool's advantages, almost all participants mentioned that the design provided by the system was useful.
All participants agreed that natural language provided an intuitive and precise means to communicate their design needs.
$\mathrm{P_2}$ mentioned, \textit{``Describing design requirements in natural language is very intuitive. 
I can simply describe what I want, and the tool would recommend proper designs.
Additionally, I am pleasantly surprised by the predefined templates and auto-completion features, which also ease the prompt formulation.''}
All participants agreed that the progressive recommendation met their expectations.
Compared to fully automated or fully manual circos plot design, participants preferred the progressive recommendation of the design.
\textit{Compared to fully automated recommendations, I can control the pace of the entire recommendation process.}-$\mathrm{P_6}$.
\textit{``progressive recommendation better aligns with my evolving design needs.''} -$\mathrm{P_4}$.
Most participants were satisfied with the recommended results.
$\mathrm{P_4}$ noted, \textit{``
The recommendation results meet my expectations and can be seamlessly applied without adjustments.
''}
As for \reference, all users acknowledged its effectiveness. 
$\mathrm{P_1}$ stated: \textit{``This presentation of track configurations allows me to intuitively overview others' designs.''}
$\mathrm{P_8}$ noted: \textit{``\reference inspire alternative designs to try, further contributing to the design process.''}
Additionally, some users felt that \reference provided some explanation for the recommendations in \recommend. 
\textit{``I notice in \reference that four papers have used the recommended results from the model, which increase my trust in those recommendations.''}-$\mathrm{P_3}$.
$\mathrm{P_2}$ suggested: \textit{``The directed acyclic graph display has a certain learning curve. 
It may be helpful to show a reference illustration from the paper directly when hovering over a specific path.''} 

G2: Simplify the implementation of \vshort.
All participants agreed that \tool streamlined the process from design to implementation of circos plots.
Most participants favored the direct application feature for recommended results.
\textit{``I can easily apply the recommended results in \recommend and \reference, which makes it easy to experiment with designs and assess their effects. 
It saves me a lot of time configuring on my own---\tool really helps me create.''}-$\mathrm{P_2}$.
Participants also agreed that the tool for supporting easy refinement of created circos plots. 
$\mathrm{P_3}$ commented \textit{``The tool provides clear options for adjusting the visual elements, making it easy to tailor the track to my needs.''}
$\mathrm{P_5}$ highlighted \tool's immediate visual feedback, stating, \textit{``
I can immediately see the results after modifying the configuration, which reduces the costs for trial-and-error.''}
However, more than half of the participants mentioned that \tool could continue to evolve to support more complex, customizable features beyond those currently available in \textbf{Circos}, such as split karyotype plots and log-scale color mappings.

\section{Discussion}
We first discuss the lessons learned from our research in Section 9.1, followed by the limitations and future work in Section 9.2.

\subsection{Design Lessons} \label{sec:lesson}
In this section, we present the design lessons learned, which can serve as inspiration for developing future tools.

\rev{\textbf{Streamlining the Creation Process from Design to Implementation.}}
\rev{
\tool simplifies the creation of circos plots by providing AI-generated recommendations and similar design references sourced from high-quality examples. 
The tool also automates the implementation of designs while offering customizable configuration options for fine-tuning the created visualization.
One important lesson we learned was balancing simplicity and flexibility in the design-to-implementation workflow. 
Our findings revealed that most participants preferred the progressive recommendation approach over fully-automated suggestions or manual design, as it allowed them to iteratively refine the results to better align with their preferences. 
This highlights the importance of workflows that adapt to users' evolving needs.
Natural language input further simplified the design process by allowing users to intuitively express their requirements. 
However, early testing revealed challenges in formulating complex prompts. 
To address this, we integrated predefined templates and auto-completion features, which were well-received by the participants.
This underscores the value of intuitive UI enhancements in reducing the learning curve.
While track-level configuration provides flexibility for customization, participants found it cumbersome when many adjustments were needed. 
A promising future direction is to enable users to select circos plots as references, allowing the tool to learn from these examples and recommend similar designs that better align with user preferences, thereby reducing the need for manual adjustments.
}

\rev{\textbf{Facilitating Effective Collaboration Between Humans and AI.}}
\rev{
\tool adopts a Human-AI collaboration workflow to assist in creating circos plots. 
Users can express their design requirements in natural language, receiving AI-generated recommendations and similar design references.
Participants in our study praised the tool in fostering creativity and inspiration, which is achieved by its AI recommendations and reference examples. 
These references not only guided their designs but also increased trust in the recommendations. 
This underscores the importance of transparency in AI assistance.
In this regard, we believe embedding contextual information, such as provenance or real-world examples, can enhance user confidence in AI-driven tools.
Furthermore, participants also appreciated the ability to combine AI recommendations with user-driven refinements, which allowed them to control the pace of design and implementation. 
This highlights the importance of enabling users to actively guide and refine AI outputs, rather than solely relying on automated suggestions.
}

\subsection{Limitations and Future Work} \label{sec:fuwork}
In this section, we discuss the limitations and future work, regarding the functionalities, and evaluations of \tool.

\textbf{Functionalities.}
The functionalities of \tool can be further extended.
The tool currently does not support sufficient data processing to prepare the data for circos plots, it only supports mapping the data to tracks.
In the future, the tool can be enhanced to support data processing tasks, \eg data filtering, so as to broaden the use cases of \tool.
Currently, users need to modify relevant configurations one by one to adjust the chart to meet their needs, which is kind of cumbersome.
In the future, \tool can be enhanced to recommend tailored configurations.
By learning from users’ created circos plots, the tool could suggest preferences for track combinations, track types, color schemes, and more, streamlining the customization process.

\textbf{Evaluations.}
In our user study, participants were tasked to \rev{create} circos plots using \tool to evaluate its support for plot creation.
However, the evaluation process could be improved from several perspectives.
First, while we assessed the usability and effectiveness of \tool's main features, we did not evaluate AI assistance in terms of trust and perceived cognitive load, which is an interesting part to dive deeper for empirical insights as suggested by \cite{yin2019understanding, gaba2023my}.
Examining these aspects would offer additional insights into the tool’s overall effectiveness.
\rev{
Second, this paper primarily focuses on the design and development of the tool, and as outlined in \autoref{sec:study design}, we did not include a baseline comparison in our study. 
However, a thorough empirical investigation comparing different approaches (e.g., \tool and Circos software) could provide a broader understanding of their respective strengths and limitations.
}
\rev{Third}, the study’s participant pool was limited.
Future work should involve a larger and more diverse group, including those not experienced creators
to better assess the tool’s real-world value.
\rev{
Lastly, all questions in the questionnaire are positive.
A better design approach would involve inverting the polarity of questions to validate participants' responses.
}

\section{Conclusion}
In this paper, we introduce \tool, an AI-powered interactive authoring tool that streamlines the design and implementation of circos plots.
Utilizing the capabilities of LLMs, \tool allows \user to specify their requirements in natural language and receive tailored recommendations to navigate the design of circos plots.
The tool then automatically applies the recommended design and supports easy refinement of the created circos plots.
Specifically, the recommendation process is enhanced by distilling commonly adopted design patterns from existing circos plots with the RAG technique.
A user study indicates the effectiveness and usability of \tool in aiding the authoring of circos plots.
Furthermore, participants highly appreciated \tool' support for the iterative creation of circos plots.

\section{Acknowledgments}
This work was supported by the National Natural Science Foundation of China (Nos.62172289).

\bibliographystyle{eg-alpha-doi} 
\bibliography{egbibsample}

\end{document}